\documentclass[aps,twocolumn,showpacs,superscriptaddress,groupedaddress]{revtex4}  
\usepackage{color}
\usepackage[usenames,dvipsnames,svgnames,table]{xcolor}
\usepackage{hyperref}
\usepackage{amsmath}
\usepackage{amsfonts}
\usepackage[utf8x]{inputenc}
\usepackage[T1]{fontenc}
\usepackage{subdepth}
\usepackage{color}

\begin{document}
\def\calr{{\cal R}}
\def\GB{{\hat{\cal{G}}}}
\newcommand{\RR}{(*R*)}
\def\half{\textstyle{1\over2}}
\def\third{\textstyle{1\over3}}
\def\quarter{\textstyle{1\over4}}
\def\nn{\nonumber}
\def\tap{{}^+\tau}
\def\tam{{}^-\tau}
\def\calR{\mathcal{R}}
\def\al{\alpha}
\def\be{\beta}
\def\ep{\epsilon}
\def\epa{\epsilon_{abcd}}
\def\emu{\epsilon^{\mu\nu\rho\sigma}}
\def\om{\omega}
\def\pd{\partial}
\title{Self-Dual Gravity and the Immirzi parameter}
\author{Javier Chagoya}
\email{j.f.chagoyasaldana@swansea.ac.uk}
 \affiliation{Department of Physics, Swansea University, Swansea, SA2 8PP, U.K.}
\author{M. Sabido}
\email{msabido@fisica.ugto.mx}
\affiliation{Departamento  de F\'{\i}sica, Universidad de Guanajuato, 
 A.P. E-143, C.P. 37150,\\ \hspace{0.8em}  Le\'on, Guanajuato, M\'exico.}
\begin{abstract}
  Working in the first order formalism of gravity, we propose an
  action that combines the self and anti-self-dual parts of the
  curvature and comprises all the diffeomorphism invariant Lagrangians
  that one can consider in this formalism.  The action that we propose
  is motivated by (A)dS gauge theories of gravity. We use this action
  to derive the (2+1)-dimensional version of the Immirzi
  parameter. Our derivation relates explicitly the Immirzi parameter
  to the existence of two classically equivalent actions for the
  description of gravity in (2+1) dimensions, namely the standard and
  exotic actions introduced by Witten in the description of (2+1)
  gravity as a gauge theory. This relation had been conjectured
previously in the literature, but not derived.

 
 \end{abstract}
\maketitle
\section{Introduction}
The standard formulation of gravity in four space-time dimensions,
General Relativity (GR), is well understood as a theory where the
dynamical degrees of freedom are directly given by {the spacetime
  metric}. However, it is also possible to work on
alternative formulations where the degrees of freedom are not
explicitly the metric components, but other fields in terms of which
the metric can be recovered. A remarkable and well known situation
where this approach have been fruitful is the development of Loop
Quantum Gravity (LQG), which was impulsed by the introduction of 
a real triad and a $SL(2,\mathbb{C})$ connection 1-form-
that replaced the metric as the dynamical field
of the theory. It was later realized that
these variables, now known as Ashtekar variables, arise in the ADM decomposition of the self-dual
Palatini Lagrangian (see, e.g., \cite{Romano:1991up}). In addition, it
was noted in \cite{barbero} that the connection used by Ashtekar can
be generalized in terms of a one parameter family of canonical
transformations applied to the variables of a SO(3)-ADM phase space
(see also \cite{immirizi,thieman}).  These new connections, known as
\emph{Ashtekar-Barbero} connections, avoid the necessity of
imposing reality conditions \emph{a-posteriori} at the expense of
working with a more complicated Hamiltonian constraint. The parameter
controlling the canonical transformations to get these variables is
dubbed the \emph{(Barbero)-Immirzi}
parameter. Although this parameter does not play a role  in physical
predictions at classical energy scales, it appears at the foundations of 
LQG. The Ashtekar-Barbero connection, together with the triad, serve
as canonical variables for the study of the Holst action
\cite{Holst:1995pc}, which is a generalization of the Palatini action
to include the Holst topological term weighted by the Immirzi
parameter. When the Immirzi parameter is equal to the imaginary unit, $i$, the self-dual
Palatini action is recovered.

Another example comes from Extended Theories of Gravity
\cite{Capozziello:2011et}, these formulations try to address the
observations that led to the dark energy/dark matter paradigm from a
different perspective, often via the addition of higher order
curvature invariants to the gravitational action. With the exception
of Lanczos-Lovelock theories \cite{1938,Lovelock:1971yv} these models
generically have equations of motion for the metric with derivatives
of order higher than two. However, in a formulation \`a la Palatini
where the metric and the curvature are considered as independent
variables, the equations of motion are of second order. The metric and
Palatini formulations coincide for the Einstein-Hilbert Lagrangian,
but for more general Lagrangians the solutions to the Palatini
formulation are a subset of the solutions to the metric equations
\cite{Exirifard:2007da,BasteroGil:2009cn}.

The alternative theory of gravity that is at the basis of this work is
known as MacDowell-Mansouri (MM) gravity \cite{PhysRevLett.38.739}
(see \cite{Blagojevic:2012bc} for a review). This theory is an attempt
to recast gravity as an ordinary Yang-Mills gauge theory, constructed
purely from the field strength of the gauge potential on either the de
Sitter (dS) group $SO(4,1)$ for positive cosmological constant, or anti-de
Sitter (AdS) $SO(3,2)$ for negative cosmological constant. This gauge
potential acts as an internal (A)dS connection whose principal feature
is that it unifies the tetrad and the spin connection used in the
Palatini formulation into a single object. This is done by associating
the translational part of the gauge connection to the tetrad and the
Lorentz part to the spin connection. For a geometrical interpretation
of MM gravity see \cite{Wise:2006sm}. By explicitly breaking the
original $SO(4,1)$ or $SO(3,2)$ gauge symmetry to its Lorentz
subgroup, $SO(3,1)$, one obtains the action for MM gravity, which
turns out to be equivalent to Einstein-Hilbert gravity with
cosmological constant, supplemented with the Euler topological
term. The MM action is thus an elegant mathematical construct with
deep connections to the infra-red and ultra-violet physics of the
spacetime: it naturally includes a cosmological constant and signals
the way to the inclusion of topological terms that modify the quantum
predictions of the theory with respect to those of pure GR
\cite{Kaul:2011va}.

In this work we investigate the relation between an extension of MM
gravity and the Immirzi parameter. Some years ago \cite{obregon}, it
was suggested that the Immirzi ambiguity has  similarities
with the $\theta$-ambiguity of Yang-Mills theories. Pursuing this idea
in the context of MM gravity and supergravity allows for an
interpretation of the Immirzi parameter along the same lines as a
$\theta$-parameter \cite{Mercuri:2010yj, Obregon:2012zz}. Here we
complement this picture by obtaining the contributions from the
$\theta$-term derived in \cite{Obregon:2012zz} directly from the
explicit breaking of gauge symmetry.

An extension of the MM theory was proposed in \cite{obni} by
constructing the action with the self-dual part of the curvature. As a
result of this extension, the $SO(3,1)$ structure of the new action is
equal to the MM action plus the Pontryagin topological term and the
Holst term, which is part of the Nieh-Yan topological term and does
not affect the classical equations of motion in the absence of
torsion.

We introduce a further generalization in the same lines as \cite{obni}:
we formulate a linear combination of two MM-like actions, one for the
self-dual part of the curvature and the other for the anti-self-dual part. Our motivation
to explore this generalization 
is not to extend the geometrical content of the theory but to gain
 freedom in the resulting coefficient for the Holst term, which
is then identified as the Barbero-Immirzi parameter. Finally, using the self-dual formulation for CS gravity, we conjecture that  the Immirzi ambiguity is related to the SL(2,{\bf Z}) invariance of the CS partition function.

The structure of the paper is as follows: in section \ref{mmtop} we
present the Lagrangians that are admissible in a first order
formulation of gravity, and we review how a subset of those terms is
recovered from the original MM action. In section \ref{asdgrav} we
review the self-dual formulation and we introduce the contributions
from the anti-self-dual curvature.  A further step is presented in
section \ref{r21} where we obtain the (2+1)-dimensional version of our
action and explore its relation to the two classically equivalent
actions for gravity in this number of dimensions.  Section \ref{con}
is devoted to conclusions.

\section{Four dimensional gravity and topological terms} \label{mmtop}
In this section we expand on the details of some topics mentioned in
the introduction, this is done in order to set up our framework and to
fix notation.  Given a theory with a certain set of symmetries and
variables, one can write down an action by adding all the terms that
can be constructed from those variables in compatibility with the
given symmetries. For four dimensional gravity in a first order
formalism we are then talking about the tetrad $e^\mu_\nu$, the
$SO(3,1)$ connection 1-form $\omega^{ab}_{\mu}$ and the groups of
diffeomorphisms and local Lorentz invariance, where Greek and Latin
letters stand for space-time and internal $SO(3,1)$ indices
respectively, both running from zero to three.  Under these
restrictions the most general\footnote{We do not consider boundary
  terms. For a study of those terms see \cite{Corichi:2016zac}.} first
order action for gravity contains the Einstein-Cartan, Holst and
cosmological constant terms, and it can be supplemented with the only
topological invariants that can be constructed in four spacetime
dimensions, namely the Euler, Pontryagin and Nieh-Yan terms (see,
e.g. \cite{Zanelli:2005sa}). All of those terms are respectively
defined below (mod.  global constant coefficients): \begin{subequations}
\label{topote}
\begin{eqnarray}
\mathcal L_{EC} &= -\ep^{\mu\nu\rho\sigma}\ep_{abcd}e^a_\mu e^b_\nu R^{cd}_{\rho\sigma} ,\label{lec}\\
\mathcal L_{H} &= -\ep^{\mu\nu\rho\sigma}e^a_\mu e^b_\nu R_{\rho\sigma a b} ,\label{lh}\\
\mathcal L_{cc} & = \emu\epa e^a_\mu e^b_\nu e^c_\rho e^d_\sigma ,\label{lcc}\\ 
\mathcal L_{E} &= \emu\epa R^{ab}_{\mu\nu} R^{cd}_{\rho\sigma} ,\label{le}\\
\mathcal L_{P} &= \emu R^{ab}_{\mu\nu} R_{ab\rho\sigma}   , \label{lp}\\
\mathcal L_{NY} & =\emu T^a_{\mu\nu} T_{\rho\sigma a} - \emu e^a_\mu e^b_\nu R_{\rho \sigma a b}, \label{lny}
\end{eqnarray}
\end{subequations}
where $\ep$ is the Levi-Civita tensor density and the curvature is
\begin{equation}
R^{ab}_{\mu\nu} = \pd_\mu \om_\nu^{ab} - \pd_\nu \om_\mu^{ab} + \om_\mu^{ca}\om_{\nu c}^b
-\om_\nu^{ca}\om_{\mu c}^b. \label{ricci4}
\end{equation} 
The Euler term is equivalent to the Gauss-Bonnet
term $R^2-4 R^{\mu\nu}R_{mu\nu} + R^{\mu\nu\rho\sigma}R_{\mu\nu\rho\sigma}$, this can be
seen explicitly by expanding the product of Levi-Civita densities in terms of
Kronecker deltas. The Pontryagin term is related to the more familiar
 Chern-Simons term which only exists in odd number of dimensions. 
The Nieh-Yan term reduces to the Holst term when the torsion vanishes.
An action containing the Lagrangian densities \eqref{lec}, \eqref{lcc} and \eqref{le} can be 
obtained from the Macdowell-Mansouri proposal,
\begin{equation} \label{mm}
S_{MM}=\int d^4 x \ \epsilon^{\mu\nu\alpha\beta}\epsilon_{abcd}\mathcal{R}^{ab}_{\mu
\nu}\mathcal{R}^{cd}_{\alpha\beta}.
\end{equation}
The  curvature  $\mathcal{R}^{ab}_{\mu\nu}$ is the four dimensional
part of the curvature associated to the five dimensional internal group and it contains
the usual curvature, eq. \eqref{ricci4}, plus some other terms related to the tetrad, i.e., it unifies
the variables of a formulation a la Palatini in a single object. 
Let us write this in detail. 

We take a connection $\omega_\mu{}^{AB}$, where as before Greek indices run
from zero to three and they label a base four-dimensional Lorentzian spacetime $\mathcal M$, while
 capital Latin letters run from zero to four and they are associated 
to a $SO(4,1)$ or $SO(3,2)$ fiber bundle attached to the base manifold. The curvature of this internal
(A)dS connection is
\begin{equation}\label{5dcurv}
\mathcal{R}^{AB}_{\mu\nu} = \partial_\mu\omega_\nu{}^{AB}-\partial_\nu\omega_\mu{}^{AB}
+\frac{1}{2}f^{[AB]}_{[[CD][EF]]}\omega_\mu{}^{CD}\omega_\nu{}^{EF},
\end{equation}
where 
 \begin{align}
 f^{[AB]}_{[[CD][EF]]}= & \eta_{AC}\delta_B^{[E}\delta_D^{F]}-\eta_{AD}\delta_B^{[E}\delta_C^{F]} \nonumber \\ 
 & +\eta_{BD}\delta_A^{[E}\delta_C^{F]}-\eta_{BC}\delta_A^{[E}\delta_D^{F]}
 \end{align}
 are the structure constants of the gauge group. The components of $f^{[AB]}_{[[CD][EF]]}$  can be separated according to the sector of the
internal space to which they belong as 
 $f^{[4a]}_{[4b][4c]}$, $f^{[4d]}_{[ab][cd]}$, $f^{[ab]}_{[4c][de]}$, $f^{[4a]}_{[4b][cd]}$, $f^{[ab]}_{[4c][4d]}$ and $f^{[ab]}_{[cd][ef]}$,
 where $a,b=0,1,2,3$. Then the five dimensional curvature \eqref{5dcurv} splits into
a purely four dimensional part, $\mathcal R^{ab}_{\mu\nu}$, and a component along the
 extra dimension, $\mathcal{R}^{a4}_{\mu\nu}$ as
\begin{subequations}
\begin{eqnarray} 
{\cal R}^{a4}_{\mu \nu} =  \partial_\mu e_{\nu}\,^a - \partial_\nu
e_{\mu}\,^a + (e_{\mu}\,^b \omega_{\nu b}\,^{a} - \omega_{\mu b}\,^a
e_\nu\,^b ), \label{torsion} \\
{\cal R}^{ab}_{\mu \nu} = R^{ab}_{\mu \nu} -  \lambda
(e_{\mu}\,^a e_{\nu}\,^b - e_{\nu}\,^a e_{\mu}\,^b)  ,\label{rfour} 
\end{eqnarray}
\end{subequations}
where $ e_{\mu}\,^a \equiv \omega_{\mu}\,^{4a} $ and $\omega_\mu{}^{ab}$ are the
 components of  $\omega_\mu{}^{AB}$ when $A$ and $B$ take values between zero and three,
and the `cosmological constant' $\lambda$ comes from $\eta^{44}$, i.e., it is given
by the internal group. $R^{ab}_{\mu\nu}$ is the usual four dimensional Riemann
curvature tensor and  ${\cal R}^{\rm a4}_{\mu \nu} $ is equal to the torsion.
 The modified four dimensional curvature \eqref{rfour} is the field
used in the MM action \eqref{mm}.

The MM action is obtained by writing a $SO(4,1)$ or $SO(3,2)$
quadratic action for $\mathcal R^{AB}_{\mu\nu}$, and explicitly
breaking the (A)dS group down to its Lorentz subgroup by tracing the
internal indices with $\epsilon_{ABCD5}\equiv\epsilon_{abcd}$, i.e.
 \begin{equation}
 S_{MM} = \int d^4 x \epsilon^{\mu\nu\alpha\beta}\epsilon_{ABCD5}\mathcal R^{AB}_{\mu\nu} \mathcal R^{CD}_{\alpha\beta}.
 \end{equation} 
This way to break the (A)dS group automatically keeps the torsion out of the action.

In the next section we present the anti-self-dual extension of \eqref{mm}, and 
we introduce the necessary ingredients to split it into the Lagrangian
 densities shown in eqs. \eqref{lec}-\eqref{lny}. Here we
skip the splitting of  action \eqref{mm} since it is along the same
lines that we follow below for the splitting of the  anti-self-dual extension.

 \section{(Anti-)self-dual gauge theory of gravity} \label{asdgrav}
 In this section we present an action that includes the
 Einstein-Cartan, cosmological constant, Euler, Holst and Pontryagin
 terms. We show that when the action combines the self-dual and
 anti-self-dual parts of the curvature, the coefficient of the Holst
 term is free and can therefore be interpreted as the Immirzi
 parameter. We closely follow \cite{obni}. As we explained in the
 previous section, the idea is to consider a five dimensional (a)dS
 internal group, whose gauge field defines a curvature. With the four
 dimensional internal part of this curvature we formulate a quadratic
 action, and we decompose this action in terms of the standard 4d
 curvature and the tetrad.

In \cite{obni}, the MM action is generalized by  considering the 
 self-dual part of the connection, given by the upper sign in 
 \begin{equation}
 ^\pm\omega_\mu{}^{ab} = \frac{1}{2}\left( \omega_\mu{}^{ab} \mp \frac{i}{2}\epsilon^{ab}_{cd}\omega_\mu{}^{cd} \right), \label{asdspin}
 \end{equation}
  to construct the corresponding quadratic action
\begin{eqnarray}\label{actsd}
S_{SD} &=& \int d^4 x \; \epsilon^{\mu\nu\alpha\beta}\epsilon_{ABCD5}{}^+\mathcal{R}^{AB}_{\mu\nu}{}^+\mathcal{R}^{CD}_{\alpha\beta}
\nonumber \\
&=& \int d^4 x \; \epsilon^{\mu\nu\alpha\beta}\epsilon_{abcd}{}^+\mathcal{R}^{ab}_{\mu\nu}{}^+\mathcal{R}^{cd}_{\alpha\beta},
\end{eqnarray}
where ${}^+\mathcal R^{ab}_{\mu\nu}$ is the self-dual part of \eqref{rfour}, given by the upper sign in
\begin{equation}
{}^\pm\mathcal R^{ab}_{\mu\nu} = \frac{1}{2}\left( \mathcal R ^{ab}_{\mu\nu} \mp \frac{i}{2}\epsilon^{ab}{}_{cd}\mathcal R^{cd}_{\mu\nu}\right). \label{asdricci}
\end{equation}
In \eqref{asdspin} and \eqref{asdricci} we already defined the anti-self-dual connection ${}^{-}\omega_{\mu}{}^{ab}$ and curvature ${}^-\mathcal R^{ab}_{\mu\nu}$
for future use. To write the forthcoming expressions in a more compact way  it is convenient to
introduce the quantities $\Sigma^{ab}_{\mu \nu} = e_{\mu}\,^a e_{\nu}\,^b - e_{\nu}\,^a
 e_{\mu}\,^b$, whose (anti-)self-duals are
 \begin{equation}
^\pm \Sigma^{ab}_{\mu \nu} = {\frac{1}{2}} (\Sigma^{ab}_{\mu \nu} \mp
 {\frac{i}{ 2}} \epsilon^{ab}\,_{cd}\, \Sigma^{cd}_{\mu \nu}).\label{sigdec} 
\end{equation}
 The (anti-)self-dual curvature is then  
 \begin{eqnarray} 
  ^\pm \calr^{ab}_{\mu\nu}&=&{}^{\pm}R^{ab}_{\mu\nu}-\lambda {}^\pm \Sigma^{ab}_{\mu\nu} \nonumber \\
 & = & \partial_\mu {}^\pm \omega_{\nu}\,^{ab} -
\partial_\nu{} ^\pm{\omega}_{\mu}\,^{ab} + ~ ^\pm{\omega}_{\mu}\,^{ca} ~
^\pm{\omega}_{\nu c}^{~~b} \nn \\
& &  - ~ ^\pm{\omega}_{\nu}\,^{ca} ~ ^\pm \omega_{\mu c}^{~~b}  -\lambda {}^\pm \Sigma^{ab}_{\mu\nu}.
  \label{rcaltorsigma}  
\end{eqnarray}
 It is convenient to note that 
 \begin{equation} 
^\pm R^{ab}_{\mu\nu}=\half\left( R^{ab}_{\mu\nu}\mp\half \imath\epsilon^{ab}{}_{cd}R^{cd}_{\mu\nu}  \right),\label{sdcdec}
\end{equation}
and from this expression it is easy to show
\begin{equation}
\epsilon^{ab}{}_{cd}{}^{\pm}R^{cd}_{\mu\nu}=\pm i {}^{\pm}R^{ab}_{\mu\nu}.
\end{equation}
 Using eqs. \eqref{sigdec}-\eqref{sdcdec} to rewrite  \eqref{actsd} in terms of $R^{ab}_{\mu
\nu}$ and $\Sigma^{ab}_{\mu\nu}$ we can see that it is equivalent to the
 MM action plus the Holst and Pontryagin terms.
 
  Now let us propose a further
 generalization by considering an action that combines the self-dual and the
 anti-self-dual parts of the connection:
\begin{align} 
S &=\mathcal G_N \int d^4 x\, \epsilon^{\mu \nu \alpha \beta}
 \epsilon_{abcd} \left( \, ^+\tau
 ^+{\cal R}^{ab}_{\mu \nu}\,  ^+{\cal R}^{cd}_{\alpha \beta} \right. \nn \\
  & \left. \hspace{4em}- \,^-\tau
 ^-{\cal R}^{ab}_{\mu \nu}\,  ^-{\cal R}^{cd}_{\alpha \beta}\right),
 \label{principalaction}
 \end{align}
where ${}^+\tau$, ${}^-\tau$ and $\mathcal G_N$ are arbitrary parameters that will at the end be
 related to the Immirzi parameter, effective cosmological constant and gravitational constant.
  Rewriting this action in terms of the
 four dimensional curvature and tetrad we obtain
\begin{align} 
 S  =&\mathcal G_N \int d^4 x\; \epsilon^{\mu \nu \alpha \beta}\,
 \epsilon_{abcd}[(\tap ^+ R^{ab}_{\mu\nu}{}^+R^{cd}_{\alpha\beta}-\tam ^-R^{ab}_{\mu\nu}{}^-R^{cd}_{\alpha\beta})
 \nn \\ &\hspace{2em} -2 \mathcal G_N  \lambda(\tap^+\Sigma^{ab}_{\mu\nu}{}^+R^{cd}_{\alpha\beta}-\tam^-\Sigma^{ab}_{\mu\nu}{}^-R^{cd}_{\alpha\beta})
 \nn \\ &\hspace{2em} +2\mathcal G_N \lambda^2 (\tap-\tam) e^a_\mu e^b_\nu e^c_\alpha e^d_\beta. \label{sasd}
\end{align}
In the last term we see the Lagrangian density $\mathcal L_{cc}$, and we can already
read off the effective value of the cosmological constant as 
\begin{equation}
\lambda_{eff} =\mathcal G_N  \lambda^2 ({}^+\tau - {}^- \tau).
\end{equation}
As a reference, note that setting $^-\tau=0$ and $^+\tau=1$, and dropping all the $^+$ superscripts what is
left is the decomposition of the MM action: it includes -- in addition to $\mathcal L_{cc}$ -- the
Einstein-Cartan and  Euler terms. 

Using \eqref{sigdec} and \eqref{sdcdec} we can expand the (anti)-self-dual curvatures to write
\eqref{sasd} in terms 
of $R^{ab}_{\mu\nu}$ and $e^a_\mu$. After a straightforward calculation we get
\begin{align}
   S = & \mathcal G_N\!  \int \!d^4 x \left[ (\tap-\tam) \epsilon^{\mu \nu \alpha \beta}\,
    \epsilon_{abcd}\left( \half R^{ab}_{\mu\nu}R^{cd}_{\alpha\beta}  \right. \right. \nn \\
     & \left. \left. \hspace{6em} - \lambda \Sigma^{ab}_{\mu\nu}R^{cd}_{\alpha\beta}+2 \lambda^2e^a_\mu e^b_\nu e^c_\alpha e^d_\beta  \right)   \right. \nn \\  &  \hspace{4em} \left. + (\tap+\tam) i  \epsilon^{\mu \nu \alpha \beta} \left(\eta_{ac}\eta_{bd}
      R^{ab}_{\mu\nu}R^{cd}_{\alpha\beta}  \right. \right. \nn \\ 
      & \hspace{6em}\left. \left. -2\lambda \Sigma^{fg}_{\mu\nu}R_{fg\alpha\beta} \right)\right]. \label{action4dtt}
\end{align}
Setting ${}^-\tau = 0$ this reduces to the same terms that are obtained from the self-dual Lagrangian \eqref{actsd}.
{A key difference when ${}^-\tau\neq0$ is that we are free to identify three independent parameters with physical significance:
\begin{subequations}
\begin{eqnarray}
G_N = \mathcal G_N (\tap-\tam), \\
\lambda_{eff} =  \mathcal G_N  \lambda^2 ({}^+\tau - {}^- \tau), \\
\beta^{-1} = \mathcal G_N(\tap+\tam), 
\end{eqnarray}
\end{subequations}
that correspond to the gravitational constant, the effective
cosmological constant and the Immirzi parameter. When ${}^-\tau = 0$
the product $\mathcal G_N {}^+\tau$ becomes a global coefficient of
the action and we cannot identify a free Immirzi parameter anymore;
indeed $\mathcal G_N {}^+\tau$ would be the coefficient of the
Einstein-Cartan Lagrangian, thus the Immirzi parameter would be
degenerated with the gravitational constant.

The second line in \eqref{action4dtt} does not contribute to the
classical equations of motion, this implies that the self-dual and
anti-self-dual actions, and any linear combination of them, are
classically equivalent -- mod. global coefficients that can be
reabsorbed in the bare values of $\lambda$ and $\mathcal G_N$. An
action similar to \eqref{action4dtt} was obtained in
\cite{Mercuri:2010yj} (see also \cite{Obregon:2012zz}) from different
arguments that led to a modification of the MM action by the
introduction of a $\theta$-term -- in analogy to a Yang-Mills theory
-- whose coefficient is related to our ${}^+\tau+{}^-\tau$.}

Up to now we have reviewed the MM action as well as a generalization
of it which gives raise to the most general first order gravitational
action that one can have in four dimensions (for vanishing
torsion). In addition, we proposed a further generalization which
allows us to have enough freedom to account for the cosmological
constant, the gravitational constant and the Immirzi parameter.  Now
we want to exploit the fact that all these terms are obtained from an
action that can be reduced to (2+1)-dimensions to identify the
equivalent of the Immirzi parameter in three dimensional gravity.
 \section{Reduction to (2+1)-dim \label{r21}}
In this section we derive an action for three dimensional gravity that consists of a linear
combination of two classically equivalent actions, known as the standard and exotic actions, introduced
by Witten in \cite{Witten:1988hc}.
The main feature of our derivation is that it allows for a straightforward identification
of the three dimensional equivalent of the Immirzi parameter. 
 There are a few proposals in the literature for what the Immirzi parameter
 might be in three dimensions, it is usually introduced  as a classical ambiguity in analogy
 to the role that it plays in four dimensions \cite{Bonzom2008,Basu2010,Barbosa2012}. Such an intuitive interpretation
coincides with the result that we present below.

{Using the (anti-)self-dual properties of the curvature, 
\begin{equation}
\epsilon^{ab}{}_{cd}{}^\pm\mathcal R^{cd}{}_{\mu\nu} = \pm 2 i \;{}^\pm \mathcal R^{ab}_{cd},
\end{equation}
 we can rewrite  \eqref{principalaction} as a total derivative
 }. To ease notation we briefly
 turn our formalism to differential forms. Then, action \eqref{principalaction} becomes
\begin{align}
S  & =   \int d^4 x\, \epsilon^{\mu \nu \alpha \beta}\,
 \epsilon_{abcd} \left( \, ^+\tau
 ^+{\cal R}^{ab}_{\mu \nu}\,  ^+{\cal R}^{cd}_{\alpha \beta}\right. \nn \\ 
 & \left. \hspace{12em} - \,^-\tau
 ^-{\cal R}^{ab}_{\mu \nu}\,  ^-{\cal R}^{cd}_{\alpha \beta}\right)  \nonumber \\ 
 & = 2\; i \tap \int {}^+\mathcal R \wedge {}^+\mathcal R  + 2\; i \tam \int {}^+\mathcal R \wedge {}^+\mathcal R \nonumber \\
& = 2\; i\tap \int (d^+\!A+^+\!A\wedge ^+\!A)\wedge(d^+\!A+^+\!A\wedge ^+\!A)  \nonumber \\ 
 & \quad + 2\; i\tam \int (d^-\!A+^-\!A\wedge ^-\!A)(d^-\!A+^-\!A\wedge ^-\!A),
\end{align}
where we set $\mathcal G_N = 1 $ and we introduced the connections $^\pm A$  defined by \footnote{Here $A$ is a $SO(3,1)$ connection
associated to the $\mathcal R^{ab}_{\mu\nu}$, which is the $SO(3,1)$ projection of the $SO(3,2)$ (or $SO(4,1)$, depending on
the sign of the cosmological constant) curvature $R^{AB}_{\mu\nu}$.}
 \begin{equation}
 ^\pm\calr = d^\pm A+{}^\pm A\wedge {}^\pm A .   \label{newRansatz}
 \end{equation}
 It is not hard to see that the integrands are just $d( {}^\pm A d{}^\pm A + {\textstyle{\frac{2}{3}}}{}^\pm A\wedge {}^\pm A\wedge {}^\pm A) $,
 so in $(2+1)$ dimensions the action \eqref{principalaction} is seen as
 a combination of the Chern-Simons actions for the (anti-)self-dual curvatures,
\begin{align}
 S_{CS}=&2\; i\tap\int ({}^+ A d{}^+\!A +{ \textstyle{\frac{2}{3}}}{}^+\!A\wedge {}^+\!A\wedge {}^+\!A) ) \nn \\ 
 &+ 2\; i\tam\int ({}^-\!A d{}^-\!A + \textstyle{\frac{2}{3}}{}^-\!A\wedge {}^- \!A\wedge {}^-\!A) ) .
\end{align}
{ This action is topological  and consistent with the fact that 2+1 gravity has no dynamical degrees of freedom.}
Turning back to the usual notation we have:
\begin{align}\label{actcsfull}
 \frac{ S_{CS}}{2i} &=\tap \int d^3x \; \epsilon^{\mu\nu\alpha}\left[  ^+\!A^{ab}_\mu\partial_\nu{}^+\!A_{\alpha a b} +\frac{2}{3}{}^+\!A^b_{\mu a}{}^+\!A^c_{\nu b}{}^+\!A^a_{\alpha c}  \right] \nn \\
   & \quad + (+ \to -),
\end{align}
where $a,b,c,d=0,1,2,3,$ $\eta_{ab}=\textrm{diag}(-1,1,1,1)$, and the complex (anti-)self-dual connections are
\begin{equation}
{}^\pm A_\mu^{ab}=\frac{1}{2}\left( A_\mu^{ab} \mp \frac{i}{2} \epsilon^{ab}{}_{cd}A_\mu^{cd} \right).
\end{equation}
Let $S_{CS}^+$ and $S_{CS}^-$ be the first and second integrals in \eqref{actcsfull},
 a straightforward calculation shows that
\begin{eqnarray}
S_{CS}^\pm = \int d^3x \; \frac{1}{2}\epsilon^{\mu\nu\alpha}\left( A_\mu^{ab}\partial_\nu A_{\alpha a b} + \frac{2}{3} A^b_{\mu a} A^c_{\nu b} A^a_{\alpha c} \right) \nn \\ 
\qquad \quad \mp \frac{i}{4}\epsilon^{\mu\nu\alpha}\epsilon^{abcd}\left( A_{\mu a b}\partial_\nu A_{\alpha c d} + \frac{2}{3} A^e_{\mu a} A_{\nu e b} A_{\alpha c d}  \right). \label{scs}
\end{eqnarray}
Then, each one of these actions contains the Chern-Simons action and the ``theta term''
 for the internal gauge group. The total action \eqref{actcsfull} will be  a linear combination of 
these two terms with coefficient $\tap-\tam$ for the  C-S  part and $\tap+\tam$ for the theta term.

{The previous actions are similar to the MM action in the sense that
  we have an internal group with one dimension more than the
  spacetime. Following the same idea, we use this extra dimension of
  the internal space to recover the information about the cosmological
  constant. }
We impose that the $SO(3,1)$ connection can be decomposed as
 $A_\mu^{ab}=(A_\mu^{AB}, A_\mu^{3A})=(\omega_\mu^{AB},\sqrt{\lambda} e^A_i)$ and
$\omega_\mu^{AB}=\epsilon^{ABC}\omega_{\mu C}$,  where now $A,B=0,1,2$.  In this way
we obtain
\begin{eqnarray}
& &{ S_{CS}}{^{\pm}}= \nn \\
& &\int_X {\frac{1}{2}} \epsilon^{\mu\nu\alpha} \left(
\omega^A_\mu(\partial_\nu \omega_{\alpha A} - \partial_\alpha \omega_{\nu A}) + {\frac{2 }{3}} \epsilon_{ABC} \omega_\mu^A \omega_\nu^B \omega_\alpha^C \right. \nn \\
  &+& \left. \lambda e^A_\mu(\partial_\nu e_{\alpha A} - \partial_\alpha e_{\mu A}) + 2 \lambda
\epsilon_{ABC}e_\mu^A e_\nu^B \omega_\alpha^C\right) \nn \\  
&\pm& i \sqrt{\lambda}\varepsilon ^{\mu\nu\alpha}\left(e_{\mu}^{A}(\partial _{\nu}\omega _{\alpha A}-\partial
_{\alpha}\omega _{\nu A})\right .\nn \\
&+&\left . \varepsilon _{ABC}e_{\mu}^{A}\omega _{\nu}^{B}\omega _{\alpha}^{C}+{
\frac{1}{3}}\lambda \varepsilon _{ABC}e_{\mu}^{A}e_{\nu}^{B}e_{\alpha}^{C}\right).
\label{finalaction}
\end{eqnarray}
The actions in the first and second rows of the last expression are respectively the exotic 
and standard actions $\tilde I$ and  $I$  \cite{Witten:1988hc}. The full action is then
\begin{equation}
S_{CS}=i(\tap+\tam)\tilde I -  2 \sqrt{\lambda} (\tap - \tam) I .\label{s3}
\end{equation}
As we saw in the previous section, the combination $\tap + \tam$ is related to the Immirzi
parameter, and the exact term corresponding to it is in the exotic action.  This is
consistent with the fact that the sum of the standard and exotic actions is a classical
ambiguity in $2+1$-dimensional gravity.  This explicit identification of the Immirzi
parameter in terms of the ambiguity for the classical description of three dimensional
gravity is a novel result of this work. {As already suggested in
  \cite{obregon,Mercuri:2010yj, Obregon:2012zz} the ``{\it Immirzi parameter}" is
  analogous to the ``{\it $\theta$-term}" in Yang-Mills theories. Furthermore in
  \cite{PhysRevD.61.085022} the transformation of the Chern-Simons gravity partition
  function was studied and it was concluded that eq.~\eqref{s3} has the form of actions
  transforming under modular transformations, and that the partition function transforms
  as
\begin{equation}
Z\left( \frac{a\tau+b}{c\tau+d}\right)=(c\tau+d)^u(c\bar{\tau}+d)^vZ(\tau),
\end{equation}
under modular transformations, relating actions with different coupling constants.
From this we conjecture that the Immirzi ambiguity in (2+1) dimensions can be related to
modular invariance of the partition function. }

Before concluding we want to comment further on the relation between the derivation of the
standard and exotic actions by Witten and the way -- introduced in
\cite{PhysRevD.61.085022} -- in which they emerge here.

 \subsection{Relation to Witten's derivation of the standard and exotic actions}
 For concreteness we do not review Witten's construction \cite{Witten:1988hc} here,
 instead we just state that it is based on a Pontryagin topological invariant for the
 group $SO(3,1)$ if the cosmological constant is positive (dS) or $SO(2,2)$ if it is
 negative (AdS), and the posterior reduction to three dimensions of this topological
 invariant. The relevant field strength is
  \begin{eqnarray}
  F_{\mu\nu}&=&P_A(\partial_\mu e_\nu ^A - \partial_\nu e_\mu ^A + \epsilon^{ABC}(\omega_{\mu B}e_{\nu C}+e_{\mu B}\omega_{\nu C}))  \\
 &+& J_A \left(\partial_\mu \omega_\nu^A - \partial_\nu\omega_\mu^A+\epsilon^{ABC}(\omega_{\mu B}\omega_{\nu C}+\lambda e_{\mu B} e_{\nu C})\right),\nonumber \label{fconlambda}
 \end{eqnarray}
 where $P$ and $J$ are the generators of translations and Lorentz transformations
 respectively, and the triad $e^A_\mu$ and connection $\omega^A_\nu$ are unified in a
 connection for (a splitting of) the groups mentioned above as $A_\mu = e^A_\mu P_A +
 \omega^A_\mu J_A$.  By inspection one can suspect that equations \eqref{torsion},
 \eqref{rfour} and \eqref{fconlambda} are related in some way. Of course the internal
 groups are different: here $A,B=0,1,2$ while in $\mathcal{R}^{ab}_{\mu \nu}$
 $a,b=0,1,2,3$. However we can show that the three dimensional projection of the internal
 group in \eqref{torsion} and \eqref{rfour} performed in the same way as was done to
 obtain \eqref{finalaction} reproduces exactly the components of $F_{\mu\nu}$.  To see
 this, we first write $\mathcal{R}^{ab}_{\mu \nu}$ in terms of its associated connection
 $A_\mu^{ab}$ used at the beginning of this section, obtaining
 \begin{equation}
 \mathcal{R}^{ab}_{\mu \nu}= \partial_\mu A_{\nu}\,^{ab} -
 \partial_\nu A_{\mu}\,^{ab} + A_{\mu}\,^{ca}
A_{\nu c}\,^b - A_{\nu}\,^{ca}
A_{\mu c}\,^b ,
\end{equation}
and then we split it by means of $A_\mu^{ab}=(A_\mu^{AB},
A_\mu^{3A})=(\omega_\mu^{AB},\sqrt{\lambda} e_\mu^A)$. We find the contribution to 3d as
\begin{subequations}
 \begin{eqnarray}
   {\cal R}^{AB}_{\mu \nu} &=& \partial_\mu A_{\nu}\,^{AB} - \partial_\nu A_{\mu}\,^{AB} + A_{\mu}\,^{CA}
A_{\nu C}\,^B\nn \\
 &-& A_{\nu}\,^{CA}A_{\mu C}\,^B 
 + A_{\mu}\,^{3A}
A_{\nu 3}\,^B - A_{\nu}\,^{3A}A_{\mu 3}\,^B \nn \\
 &  = &\partial_\mu \omega_{\nu}\,^{AB} -
 \partial_\nu \omega_{\mu}\,^{AB}+ \omega_{\mu}\,^{CA}
\omega_{\nu C}\,^B \nn \\
&-& \omega_{\nu}\,^{CA}\omega_{\mu C}\,^B +\lambda \left(e_\mu^A e_\nu^B -  e_\nu^A e_\mu^B\right). \label{3dcurv} \\
{\cal R}^{ A4}_{\mu \nu}& = &  \partial_\mu e_{\nu}\,^A - \partial_\nu
e_{\mu}\,^A +(e_{\mu}\,^B \omega_{\nu B}\,^{A} - \omega_{\mu B}\,^A
e_\nu\,^B ).\nn\\\label{3dtor}
 \end{eqnarray}
 \end{subequations}
The last line follows straightforwardly from  \eqref{torsion} since $e_\mu^3$ does not exist.
To compare directly with \eqref{fconlambda} we  need to write both connections in the same notation (remember that before we introduced  $\omega_\mu^{AB}=\epsilon^{ABC}\omega_{\mu C}$). It is easier to do this starting from the field strength \eqref{fconlambda} and reinserting the definition  $J^A=\half\epsilon^{ABC}J_{BC}$ \cite{Witten:1988hc} in order to form the appropriate combinations of the Levi-Civita tensor density and the spin connection:
 \begin{widetext}
 \begin{eqnarray}
   F_{\mu\nu}& =&   P_A(\partial_\mu e_\nu ^A - \partial_\nu e_\mu ^A + \epsilon^{APQ}(\omega_{\mu P}e_{\nu Q}+e_{\mu P}\omega_{\nu Q})) + J_A \left(\partial_\mu \omega_\nu^A - \partial_\nu\omega_\mu^A+\epsilon^{APQ}(\omega_{\mu P}\omega_{\nu Q}+\lambda e_{\mu P} e_{\nu Q})\right)  \\
 \phantom{ F_{\mu\nu}} &=&  P_A(\partial_\mu e_\nu ^A - \partial_\nu e_\mu ^A + \epsilon^{APQ}(\omega_{\mu P}e_{\nu Q}+e_{\mu P}\omega_{\nu Q})) +
 \frac{1}{2}\epsilon^{ABC}J_{BC}\left(\partial_\mu \omega_{\nu A} - \partial_\nu\omega_{\mu A}+\epsilon_{APQ}(\omega_{\mu}^{ P}\omega_{\nu}^{ Q}+\lambda e_{\mu}^{ P} e_{\nu}^{ Q})\right)\nn\\
     \phantom{F_{\mu\nu}}  
     &=& P_A(\partial_\mu e_\nu ^A - \partial_\nu e_\mu ^A -(\omega_{\mu}^{Q A}e_{\nu Q}-e_{\mu P}\omega_{\nu}^{P A}))+  \frac{1}{2} J_{BC}\left(\partial_\mu \omega_{\nu}^{ BC} - \partial_\nu\omega_{\mu}^{ B C} + \omega_\mu^{D B}\omega_{\nu D}{}^{ C} -\omega_\nu^{D B}\omega_{\mu D}{}^{ C}+\lambda (e_{\mu}^{ B} e_{\nu}^{ C}-e_{\mu}^{ C} e_{\nu}^{ B})\right),\nn
  \end{eqnarray}
  \end{widetext}
 where we have used 
  \begin{equation}
  \omega_\mu^D{}_B\omega_{\nu D C} - \omega_\nu^D{}_B\omega_{\mu D C}=\omega_{\mu B}\omega_{\nu C}-\omega_{\mu C}\omega_{\nu B} .
  \end{equation}
In the last expression we see  that the spacetime components of $F_{\mu\nu}$ are a linear 
combination  of the components of
\eqref{3dcurv} and \eqref{3dtor}.
This clarifies the relation between the field strength
used by Witten and the curvature that we used to identify the 3-dimensional
analogue of the Immirzi parameter.

\section{Conclusions} \label{con}
In this paper we proposed a first order action for four dimensional gravity constructed
with the self and anti-self-dual parts of the field strength of an (A)dS internal
connection.  After explicit symmetry breaking, the action reduces to a combination of the
Einstein-Cartan, Holst, cosmological constant and Euler terms. The geometrical content of
our model is the same as in \cite{obni}, where only the self-dual part of the gauge
connection is considered.

The advantage of adding the anti-self-dual terms is that now we are able to identify the
Immirzi parameter as a combination of the free parameters of the action. This combination
is not degenerated with the gravitational and cosmological constants, and it does not
appear in the classical equations of motion. The decomposed form of the action,
\eqref{action4dtt}, can be also interpreted as coming from an usual MM formulation
supplemented with a ``{\it $\theta$-term}'', as argued in \cite{Mercuri:2010yj,
  Obregon:2012zz}.

Due to the (anti)-self-duality of the fields that we use, we can write their quadratic
actions as Pontryagin terms. In consequence we can obtain their three-dimensional versions
as Chern-Simons terms. {In doing so, there is the underlying assumption that the spacetime
  has a well-defined boundary so that we can apply Stoke's theorem. Evidently, this would
  not be the case for AdS or flat spacetimes. Nonetheless, we can start with a four
  dimensional manifold that has a well-defined tree dimensional boundary where the
  combination ${}^+\tau + {}^-\tau$  is related to the four dimensional Immirzi
  parameter. We remark that our claim is not that we find a three dimensional version of
  the same physical system described by \eqref{action4dtt}, but rather our claim is as
  follows: in four spacetime dimensions we first identify a combination of parameters that
  corresponds to the Immirzi parameter, and then we take advantage of the structure of the
  action to find a situation where we can integrate out one spacetime dimension, and
  investigate the role of such combination of parameters in the resulting lower dimensional
  setting.  By a convenient splitting of the connections, the Chern-Simons form gives
  raise to the standard and exotic actions for three dimensional gravity, these actions
  are classically equivalent in the sense that upon variation they lead to the same
  equations of motion.
 Thus, what we obtain from \eqref{principalaction} in three dimensions
 is a linear combination of the standard and exotic terms. The combination of $\tap$ and $\tam$
associated to the Immirzi parameter in four dimensions
translates to the coefficient of the exotic action in the three dimensional
 action. This is satisfying from an heuristic point of view:
both in three and four dimensional space-time the Immirzi parameter plays the
role of a classical ambiguity. This identification has been proposed and used
in other works \cite{Bonzom2008,Basu2010,Barbosa2012} but without a formal
 derivation from a four dimensional action. 
 
{Finally, we conjecture that the Immirzi ambiguity can be understood from the modular invariance transformation of the (2+1)
 partition function}.
Self-dual Chern-Simons theories in three dimensions have been widely studied in
the literature for different reasons, in the context of gravity this is
generically in the search of guiding principles to construct four dimensional
theories (see for example \cite{Carlip2008,Dijkgraaf:2004te,Kuzenko2014a} for
 specific applications or \cite{Zanelli2012} for a 
general review). The analysis of what those developments can tell us about the
Immirzi parameter is a natural perspective  for this work.
\section*{Acknowledgments}
This work is  supported by   CONACYT grants 167335, 179208, 257919, 290649 and DAIP1107/2016. This work is part of the PROMEP research network  ``Gravitaci\'on y F\'isica Matem\'atica".

 \end{document}